\documentclass[11pt,a4paper]{article}

\usepackage{geometry}
\usepackage[applemac]{ inputenc}

\usepackage[british,UKenglish,USenglish,english,american]{babel}

\usepackage{mathrsfs}
\usepackage{amsmath}
\usepackage{amssymb}
\usepackage{multicol}
\usepackage{float}
\usepackage{makeidx}
\usepackage{layout}
\usepackage{array}
\usepackage{boxedminipage}
\usepackage{hyperref}
\usepackage{latexsym}
\usepackage{indent first}

\usepackage{subcaption,graphicx}
\usepackage{appendix}

\usepackage{color}

\usepackage[sectionbib,square]{natbib}

\usepackage[normalem]{ulem}

\DeclareMathOperator{\sech}{sech}

\title{Trapped solitary waves and collisions for the forced Korteweg-de Vries equation}
\author{M. V. Flamarion$^{1}$, P. A. Milewski$^{2}$ and R. Ribeiro-Junior$^{3}$}

\date{}

\begin{document}

\maketitle

{\footnotesize
	\begin{center}
	$^{1}$ UFRPE/Rural Federal University of Pernambuco, UACSA/Unidade Acad{\^e}mica do Cabo de Santo Agostinho, BR 101 Sul, 5225, 54503-900, Ponte dos Carvalhos, Cabo de Santo Agostinho, Pernambuco, Brazil.
	
	$^{2}$ University of Bath, Department of Mathematics, Bath, Somerset, United Kingdom.
	
	$^3$ UFPR/Federal University of Paran\'a,  Departamento de Matem\'atica, Centro Polit\'ecnico, Jardim das Am\'ericas, Caixa Postal 19081, Curitiba, PR, 81531-980, Brazil. 
	
	\end{center}
	
	}
\begin{abstract}
The aim of this work is to study trapped waves and their collisions between two topographic obstacles for the forced Korteweg-de Vries equation.  Numerical simulations show that solitary waves remain trapped bouncing back and forth between the obstacles until the momentum overcomes a certain threshold. We find that the waves have a certain tendency of escaping out upstream.  Furthermore, the time of escape of the trapped wave varies linearly with the distance between the bumps. 
Besides, we  study collisions of  solitary waves between the obstacles.  Although the dynamic of one wave is affected by the other one,  statistically this feature is not  evident.

 
\end{abstract}

\section{Introduction}

The forced Korteweg-de Vries equation (fKdV)  has been broadly used  to study current-wave-topography interactions (\cite{Wu1, Camassa1, Camassa2, Grimshaw94, Paul}). In this scenario, an interesting phenomenon that arises is waves that remain trapped in a certain region of the space. These waves are called trapped waves.

A remarkable work on this topic was done by \cite{Grimshaw94}. These authors  used the fKdV equation to investigate asymptotically and numerically the interaction of a solitary wave with a external force modeled by a single obstacle of small amplitude. They classified the waves solutions as: passage, when the solitary  wave passes over the bump without reversing its movement, repulsion, when the wave is reflected by the bump, and trapping, when the solitary wave oscillates back and forth over the bump. More recently, \cite{Ermakov} revisited     the asymptotically approach developed by \cite{Grimshaw94}  and studied the interaction of a solitary wave with different types of single bumps. In the Euler equations framework \cite{Flamarion-Ribeiro} investigated trapped waves in a low-pressure region. 





Considering a bottom topography with two bumps, \cite{Lee1} and \cite{Lee2} investigated solutions for the fKdV which remained oscillating back and forth between the two obstacles for a certain period of time.  Along the same lines \cite{Kim} investigated the dynamic of solitary trapped waves between two bumps and verified that the center of mass of the trapped waves has to cross an  energy barrier in order to pass over  the bumps.  They also showed that the energy barrier does not depend on the distance between the obstacles, while it increases as the height of the bumps gets higher. Regarding the time of escape of the trapped waves, they found evidences that it varies linearly with the distance between the topographic obstacles. These authors obtained trapped waves as perturbations of  steady waves solutions  for the fKdV equation. Hence these previous studies are limited to analyse  just the  case in which only one wave remain trapped. 


In this paper we study trapped wave solutions between two bumps for the fKdV equation. Differently from previous works, we consider a $\sech^{2}$-solitary wave like as an initial data and investigate regimes in which the solution oscillates back and forth between the obstacles and analyse how the time of escape varies with the distance between the bumps. Our experiments show that these two quantities are linearly dependent. Moreover, we notice that as the wave moves  back and forth between the bumps its momentum oscillates increasing until it overcomes a certain threshold and escapes out. It is remarkable that such behaviour resembles the concept of circulating power in laser light. When the light circulates back and forth between two mirrors stimulating  more and more energy on each pass until the energy reaches a certain threshold point and the light passes through one of the mirrors (\cite{Hitz}). Furthermore, we find that the trapped waves have the tendency of espaping out upstream.  We also investigate collisions of two trapped waves between the bumps and notice  that although one wave affects the other when   the problem is observed statistically the waves behave as they were almost independent. In addition, we display a regime in which three solitary waves  are colliding between the bumps and after a series of collisions  just one of them remain trapped.


The paper is organized as follows. In section 2 we present the mathematical formulation of the non-dimentional fKdV equation.  The results are presented in section 3 and the conclusion in section 4.

\section{The forced Korteweg-de Vries equation}

We consider a two-dimensional flow of an inviscid, incompressible  fluid of constant density in a  shallow water channel with uneven bottom ($h(x)$). Assuming that there is a constant current in the flow, the Froude number ($F$) is defined by the ratio of the upstream velocity and the linear long-wave speed. In the nearly-critical regime ($F\approx 1$) with obstacles of  small amplitude, the free surface displacement ($\zeta(x,t)$) is governed by forced Korteweg-de Vries equation (\cite{Wu1, Paul, Marcelo-Paul-Andre})
\begin{equation}\label{fKdV}
	\zeta_{t}+f\zeta_{x}-\frac{3}{2}\zeta\zeta_{x}-\frac{1}{6}\zeta_{xxx}=\frac{1}{2}\epsilon h_{x}(x), 
\end{equation}
where $f$ denotes a small variation  of the Froude number, i.e, $F=1+\epsilon f$, where $\epsilon>0$. 

Besides,  the equation (\ref{fKdV})  yields the  momentum $P(t)$,
\begin{equation}\label{momento}
	P(t) = \int_{-\infty}^{+\infty}\zeta^{2}(x,t)\, dx,
\end{equation}
which its  rate of change is balanced by 
\begin{equation}
	\frac{dP}{dt}=\int_{-\infty}^{+\infty}\zeta(x,t)h_{x}(x)\, dx.
\end{equation}




The traveling solitary wave  
\begin{equation}\label{solitary}
\zeta(x,t)=A\sech^{2}(k(x-ct)),  \;\ A=\frac{4}{3}k^{2}, \;\ c =f-\frac{1}{2}A.
\end{equation}
is solution of  (\ref{fKdV}) when the bottom is flat ($h_x=0$). Note that when $f=A/2$ the solution is stationary.


The solutions of the fKdV equation (\ref{fKdV}) are computed using the standard  pseudospectral numerical method shown in (\cite{Trefethen,Shen}). Moreover, we consider a periodic  computational domain which is taken large enough to avoid the effects of the spatial periodicity.




\section{Results}

\subsection{Trapped solitary waves}

We are interested in studying trapped solitary waves between two bumps. For this purpose, we solve the fKdV equation (\ref{fKdV}) with initial condition 
\begin{equation}\label{initialdata}
\zeta_0(x)=A\sech^{2}(kx), \quad \mbox{with} \;\  A=\frac{4}{3}k^{2},
\end{equation}
and topography $$h(x)=\Big[\exp{\big(-(x-\beta)^2\big)}+\exp{\big(-(x+\beta)^2\big)}\Big].$$
\begin{figure}[h!]
	\centering	
	\includegraphics[scale =1]{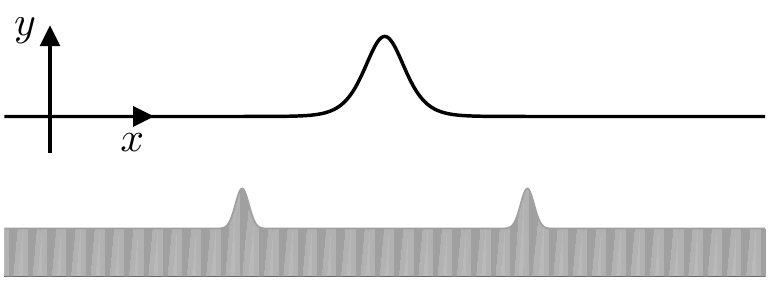}
	\caption{Sketch of the physical problem.}
	\label{sketch}
\end{figure}
In the following simulations we fix $\epsilon=0.01$.  The sketch of the physical problem is depicted in Figure \ref{sketch}.

As mentioned in the previous section, in the absence of a variable topography the choice $f=A/2$ yields a  stationary solution. Therefore, we seek for trapped waves when $f$ is close to $A/2$. 

\begin{figure}[h!]
	\centering	
	\includegraphics[width = 0.9\linewidth]{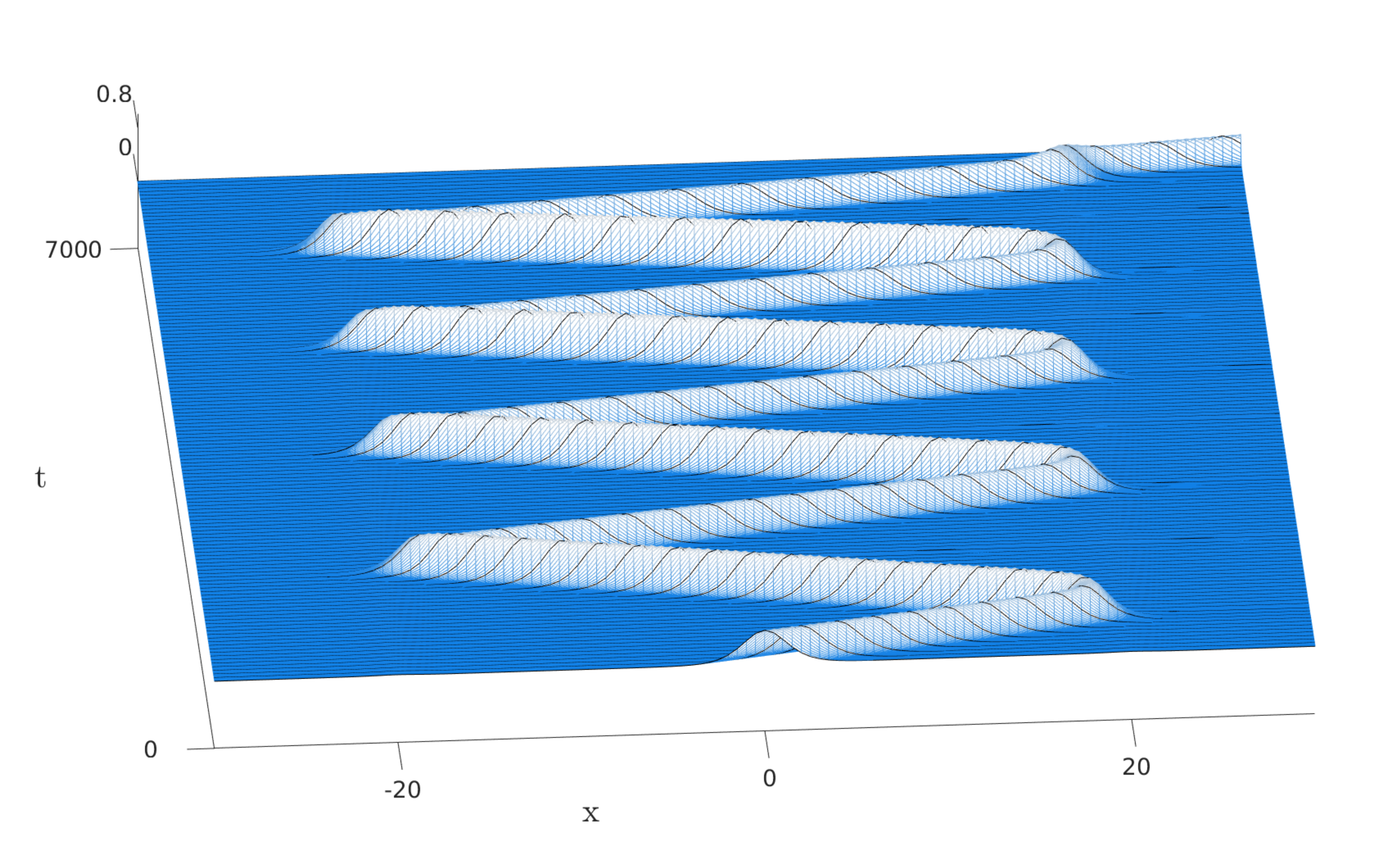}
	\caption{ Trapped solitary wave between two bumps. }
	\label{OndaPresa}
\end{figure}

Figure \ref{OndaPresa}  displays the evolution of $\zeta_0(x)$ for $A=0.5$, $f=0.29$, and $\beta=20$. The solitary wave that initially is stationed over the flat part of the topography moves to the right as a traveling wave. When it reaches the obstacle it reflects back oscillating back and forth within the two bumps. 

\begin{figure}[h!]
	\centering	
	\includegraphics[scale =1]{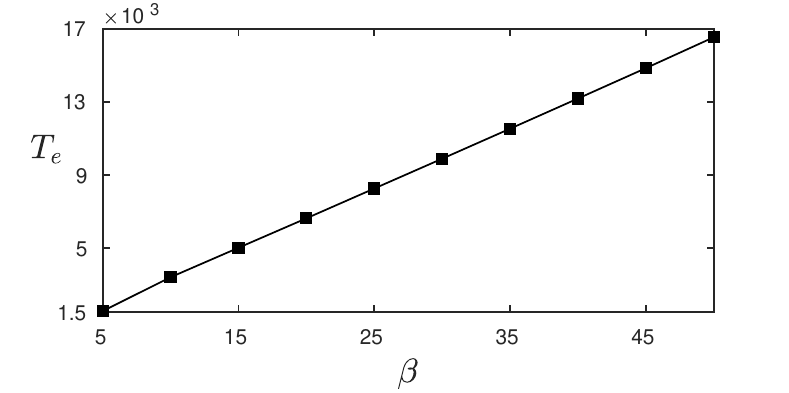}
	\caption{The time of escape ($T_e$) of the trapped wave as a function of $\beta$.}
	\label{BetaVariando}
\end{figure}

In order to investigate the trapped wave and its different nuances we fix the parameter $f$. Thus, unless mentioned otherwise $f=0.29$. With this in mind we also fix $A=0.5$ and vary the distance between the obstacles. So that, we can analyse how the topography affects the trapped wave. In our numerical simulations we have noticed that after sufficient time has elapsed the trapped waves overcome the obstacles. We say that the trapped wave escape out of the two bumps at  time $t=T_{e}$ when the highest peak of the wave reaches the positions $x=\pm\beta$.
Figure \ref{BetaVariando} shows that  $\beta$  and $T_{e}$  are linearly dependent.  Investigating the stability of steady waves for the fKdV equation \cite{Kim} found evidences that the time of escape of the trapped wave varies linearly with the distance of the two bumps. Here, we confirm their prediction for solitary-wave like.

%

\begin{figure}[h!]
	\centering	
	\includegraphics[scale =1]{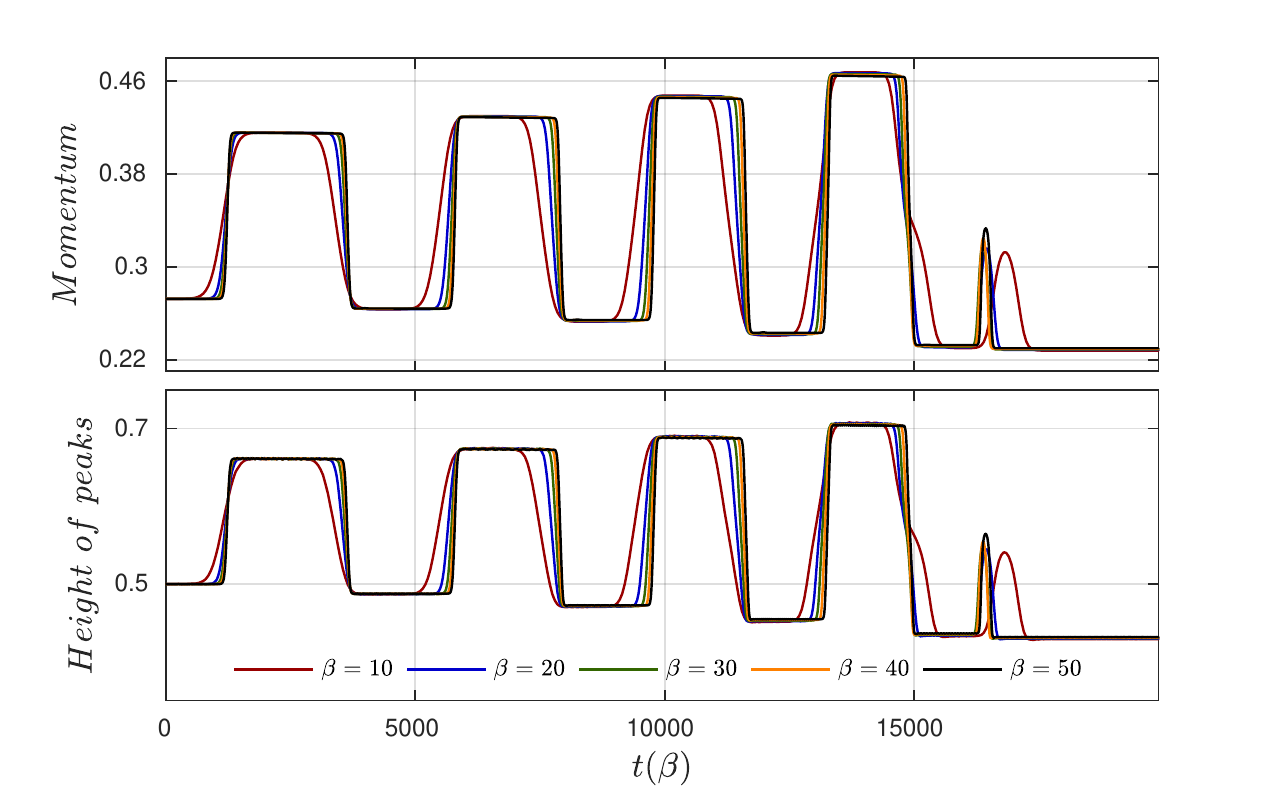}
	\caption{Momentum (top) and height of the highest peak of the wave (bottom) as a function of $t(\beta)=50t/\beta$.}
	\label{Momento}
\end{figure}

\begin{figure}[h!]
	\centering	
	\includegraphics[scale =1]{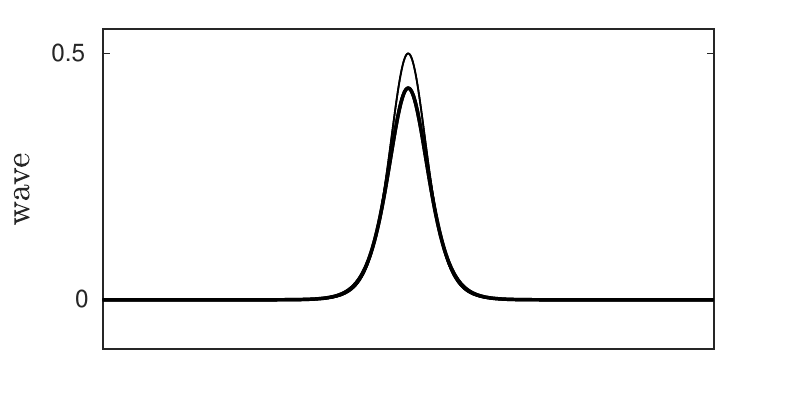}
	\caption{The dark line displays the solitary wave used as initial data. The thicker line, which is the graph of $\displaystyle{\zeta_{\star}(x)=0.43\sech^{2}(0.57x)}$, represents the behaviour of the initial data as $t$ goes to infinity for all values of $\beta$ in figure \ref{Momento}.}
	\label{OndaLimite}
\end{figure}

We are interested in determining a certain criteria that the trapped waves have to satisfy to overcome the bumps, so as before we fix $A=0.5$ and analyse the momentum (\ref{momento}) for different values of $\beta$. In our simulations we observed that the trapped wave always rebounds nine times between the obstacles before escaping out. Figure \ref{Momento}  shows the momentum (top) and the height of the highest peak of the wave (bottom) as a function of $t(\beta)$, where $t(\beta)=50t/\beta$ for $\beta=10,20,30,40,50.$ Notice that the momentum oscillates  increasing until the wave overcomes the bumps and the same goes to the height of the peak of the wave. Besides,  when the wave is traveling downstream (upstream)  its amplitude increases (decreases) as it reaches the obstacle, as if the wave was colliding against a wall. Furthermore, it is worthy of note that the trapped wave behaves like a traveling wave for certain periods of time while it is traveling between the obstacles. 
Moreover, from our numerical simulations, we observed that as the wave escapes out and moves away from the bumps its profile shape $\zeta$ approaches $\zeta_{\star}$ as $t\to \infty$, where
\begin{equation}\label{solitaryinfinity}
\zeta_{\star}(x)=A_{\star}\sech^{2}(k_{\star}x),  \;\ A_{\star}=\frac{4}{3}k_{\star}^{2}, \;\ \mbox{and}
  \;\ A_{\star}=0.430.
\end{equation}
Figure \ref{OndaLimite} depicts the profile of the initial data and $\displaystyle{\zeta_{\star}(x)}$. We point out that at least for the values of $\beta$ which we used in our simulations the limit (\ref{solitaryinfinity}) does not depend on $\beta$. 

\begin{figure}[h!]
	\centering	
	\includegraphics[scale =1]{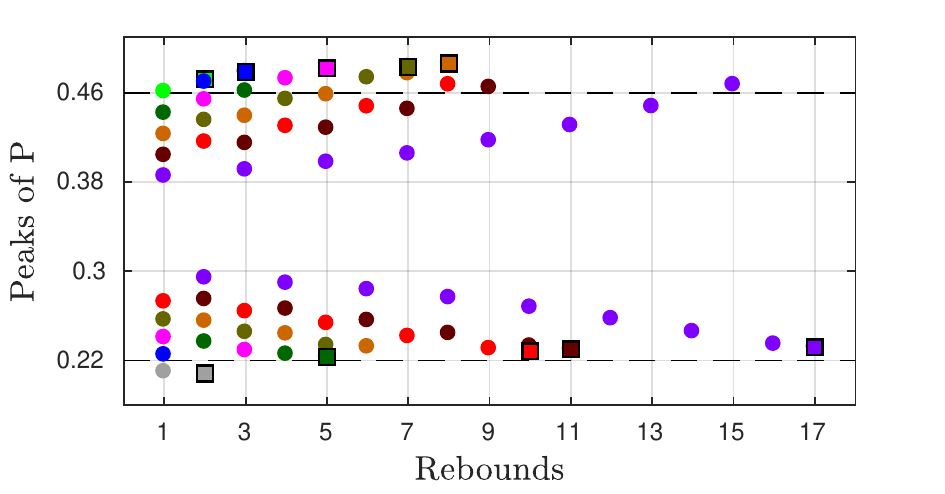}
	\includegraphics[scale =1]{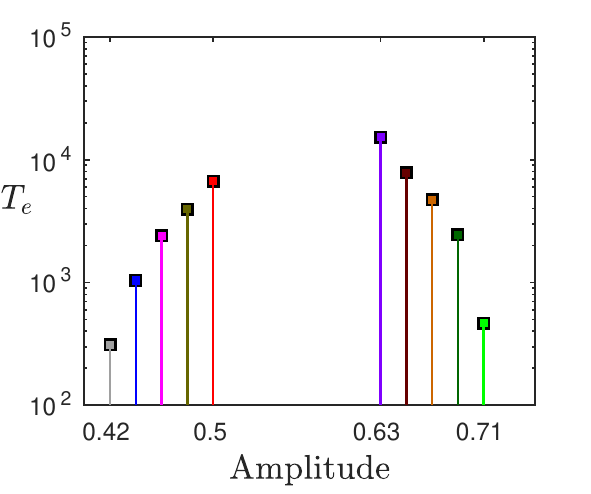}
	\caption{Left: The colored dots represent different waves and their peaks of momentum at each rebound. The colored squares display the momentum after the waves escape out. Right: The time of escape of each wave depicted on the left figure.}
	\label{colorida}
\end{figure}

In the previous simulation we fixed the amplitude of the initial data and varied the distance between the bumps. This led us to conclude that the maximum of the momentum in each oscillation does not depend on the distance between the obstacles. Now, we fix the distance between the obstacles ($\beta=20$) and vary the amplitude of the initial solitary wave. As a result of that we found that  there is an upper and lower bound for the momentum that once the wave crosses this threshold it escapes out. This is shown in Figure \ref{colorida} (left) which displays the peaks of the momentum in each rebound for different initial datas. The colored dots represent different waves and their peaks of momentum at each rebound. The colored squares display the momentum after the waves escape out. For instance, the red dots correspond to the wave whose momentum is depicted in Figure \ref{Momento}. Notice that all trapped waves cross the upper dashed line before escaping out. Besides, when the momentum of the initial wave is above the upper dashed line or bellow the lower dashed line it never rebounds (see gray and light green dots). Figure \ref{colorida} (right) shows the time of escape of each wave depicted in Figure \ref{colorida} (left).

\begin{figure}[h!]
	\centering	
	\includegraphics[scale =1]{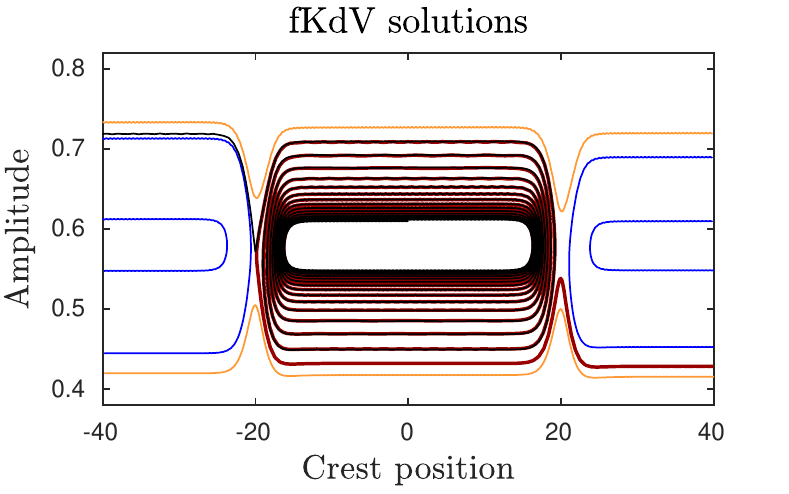}
	\caption{Amplitude of different solutions of the fKdV equation as a function of their crests position.}
	\label{EspacoSolucao}
\end{figure}

Lastly, we present Figure \ref{EspacoSolucao} which describes  several solutions of the fKdV (\ref{fKdV}) in the  Crest Position of the solitary wave vs. Amplitude plane. Although we consider two bumps the solutions of (\ref{fKdV}) can be classified in  three types as reported by \cite{Grimshaw94} for one-bump topography: passage (yellow curves), repulsion (blue curves) and trapping (black and red curves).




\subsection{Trapped wave collisions}

Now we investigate trapped wave collisions. To this end, we consider two well separated solitary waves as initial data. More precisely, we solve  (\ref{fKdV}) with initial condition
$$\zeta(x,0)= \zeta_A(x) + \zeta_B(x),$$
where 
\begin{equation}\label{dadoInicial}
\zeta_A(x)=A\sech^{2}(k_{A}(x+10)) \quad \text{ and } \quad \zeta_B(x) = B\sech^{2}(k_{B}(x-10)), 
\end{equation}
and the topographic obstacle is the same used in previous section with $\beta = 20$ and $\epsilon = 0.01$.

\begin{figure}[h!]
	\centering	
		\includegraphics[scale =1]{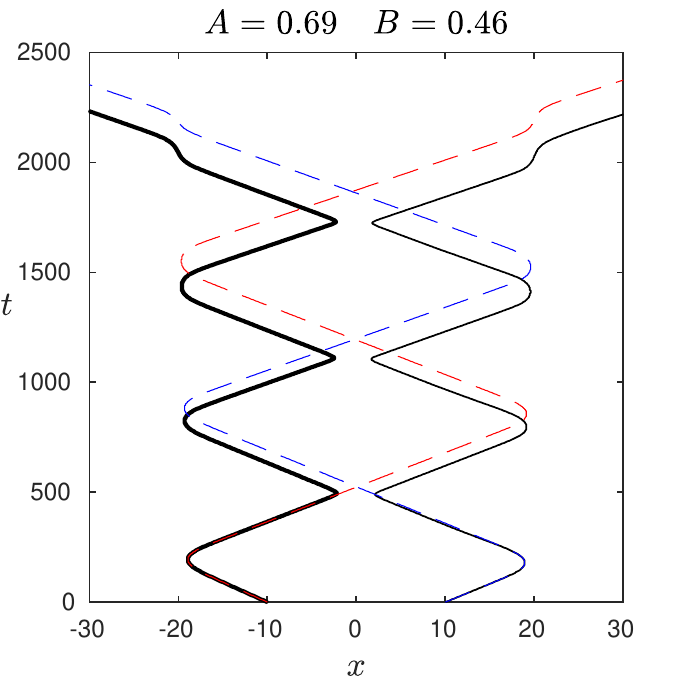}
		\includegraphics[scale =1]{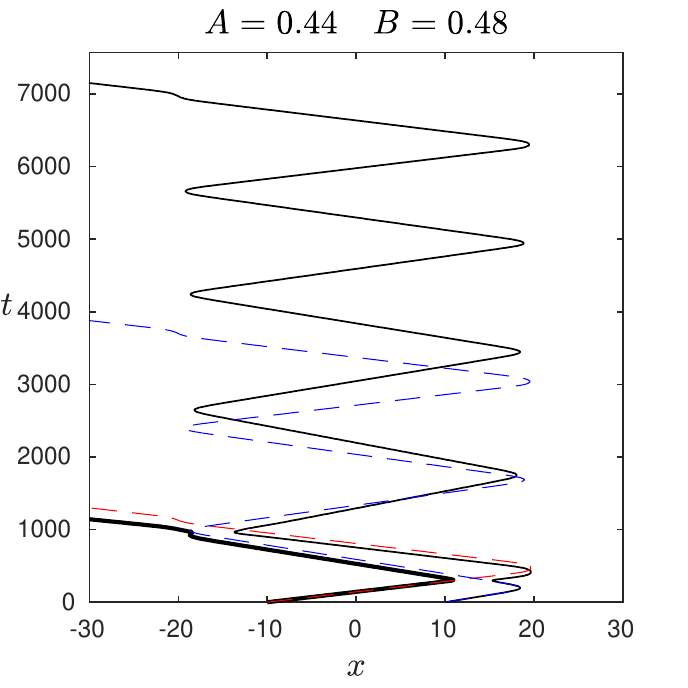}

		\includegraphics[scale =1]{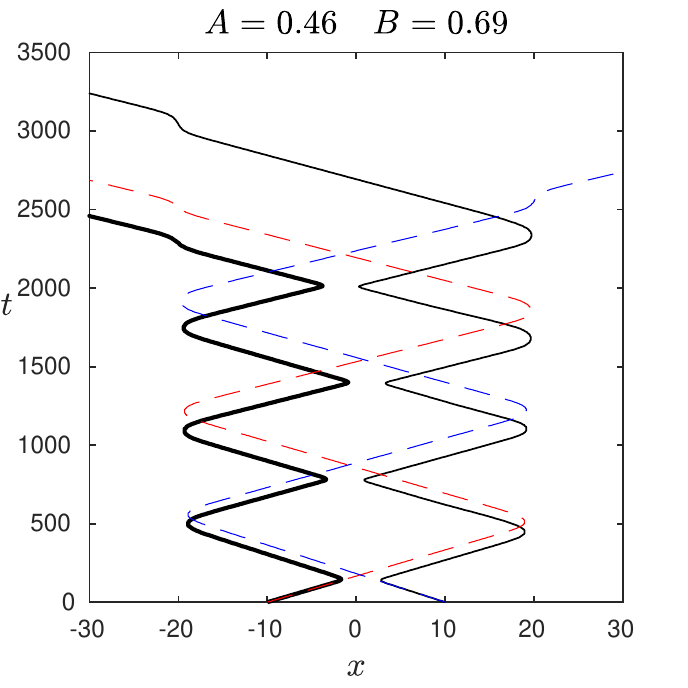}
			\includegraphics[scale =1]{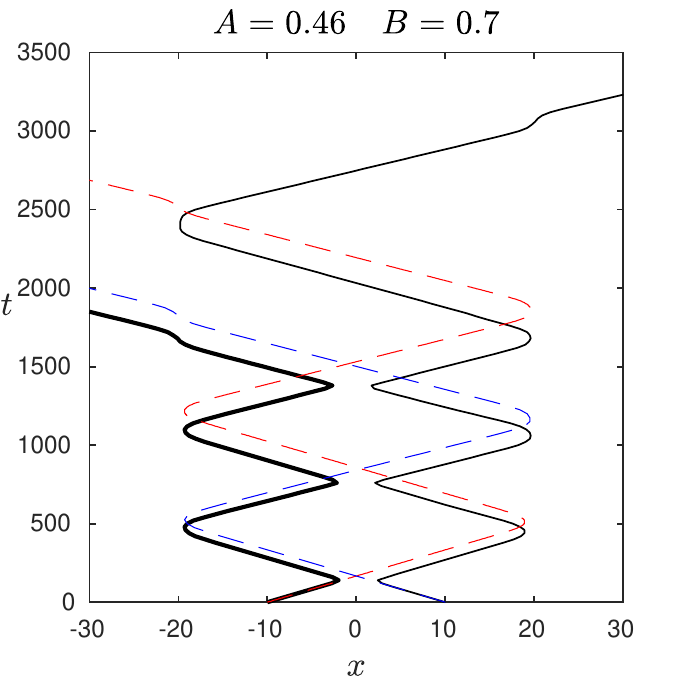}

		\includegraphics[scale =1]{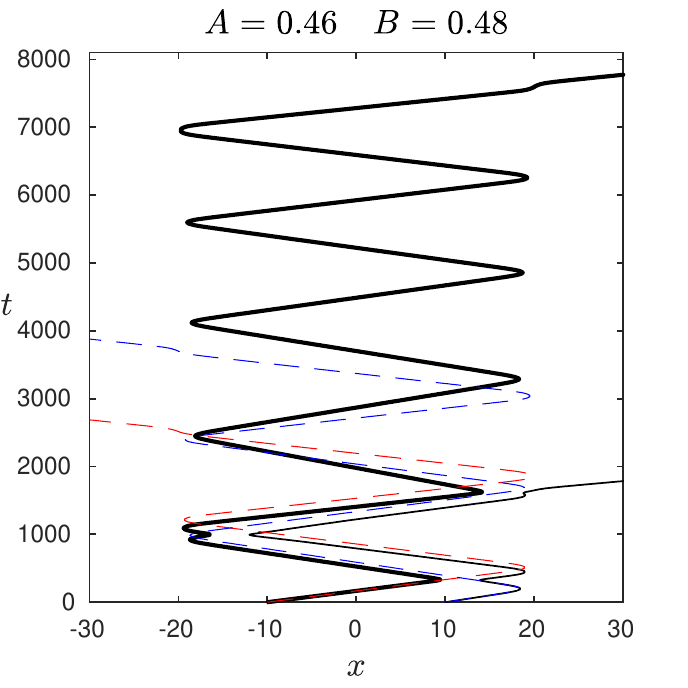}
	\includegraphics[scale =1]{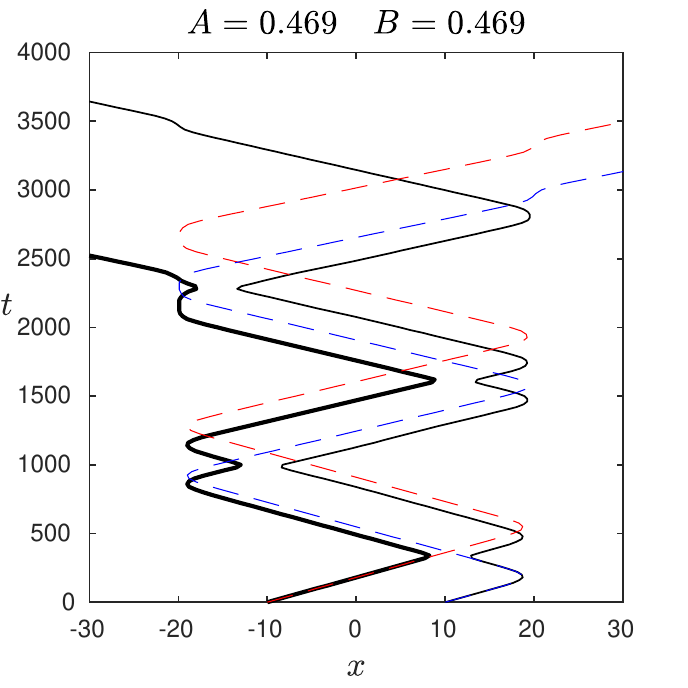}

	\caption{Crest trajectory for  three different initial conditions: $ \zeta_A$ (red), $ \zeta_B$ (blue) and $ \zeta_A + \zeta_B$ (black).		
		}	
	\label{colisao}
\end{figure}

It is well known that in the KdV model collisions of solitary waves results   only in a phase shift. In Figure \ref{colisao},  we plot, in each panel, the crest trajectories for three different initial conditions: $\zeta(x,0) = \zeta_A(x)$ (red), $\zeta(x,0) = \zeta_B(x)$ (blue) and $\zeta(x,0) = \zeta_A(x)+ \zeta_B(x)$ (black). We observe that in some cases the wave interactions cause only a phase shift, while in others the presence of a second wave affects the direction of escaping.

\begin{figure}[h!]	
	\centering
	\includegraphics[scale =1]{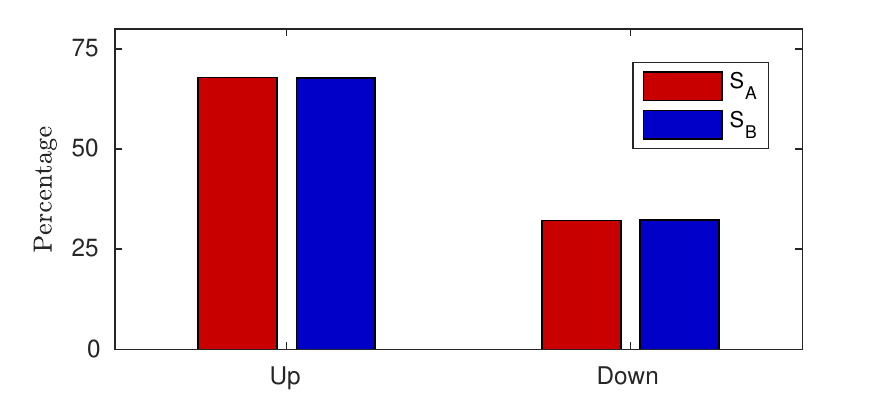}
	\caption{ The  percentage of waves that espapes out upstream in the sample $S_A$ and $S_B$ are  $67.85\%$ and $67.79\%$ respectively.}
	\label{FigPorcentagemAB}
\end{figure}

These results lead us to investigate in more details the wave  interactions 
described above. Thus, as before we solve (\ref{fKdV}) with the same initial conditions.  However, now we carry out a larger number of experiments, namely,  we consider  $A \in S_A$ and $B \in S_B$, where  $S_A$ and $S_B$ are disjoint subsets of $S= [0.437,0.52]\cup[0.64,0.709]$ with around $1500$ elements each. We point out that the choice of the set $S$ is made in order to guarantee that there is no effect of the spatial periodicity,  such as the return of small amplitude radiation  entering in neighbourhood of the obstacles.
Figure \ref{FigPorcentagemAB} displays  the percentage of escaping direction  when a single solitary wave ($\zeta_A$ or $\zeta_B$) is taken as initial data of  (\ref{fKdV}). As can be seen the wave has a tendency to escape out upstream.

\begin{figure}[h!]	
	\centering
	\includegraphics[scale =1]{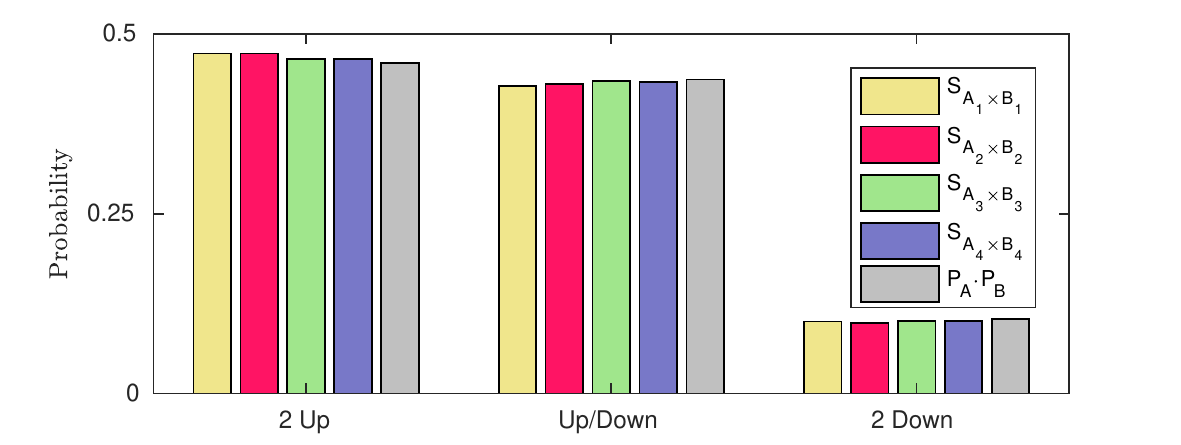}
	\caption{Probability of the outcome of two solitary wave colision.}
	\label{FigABTresEventos}
\end{figure}

 When two well separated solitary  waves ($\zeta_A + \zeta_B$) are considered as initial condition of (\ref{fKdV}) we choose  $(A,B) \in   S_A\times S_B$. In this case the dynamic can be classified as: both waves escape out upstream (2 Up), both waves espape out downstream (2 Down) or the waves escape out in opposite direction (Up/Down).  Figure \ref{FigABTresEventos} shows the probability of escaping for $(A,B) \in S_{A_i}\times S_{B_i} \subset S_A\times S_B$, $ i = 1,2,3,4$, where each subset has roughly 11,000 elements. In this figure $P_A \cdot P_B$ represents the probability of the dynamic if the presence of  one  wave would not affect the other.   Two facts are evidents. First, as the subsets $S_{A_i}\times S_{B_i}$ have almost the same probability we infer that the size of the samples is large enough to capture the features of the phenomenon. Second, although individually each wave affects the other, when we see the collisions statistically it seems as if the events were almost independent (see gray column in Figure \ref{FigABTresEventos}).



		

\begin{figure}[h!]	
	\centering
	\includegraphics[scale =1]{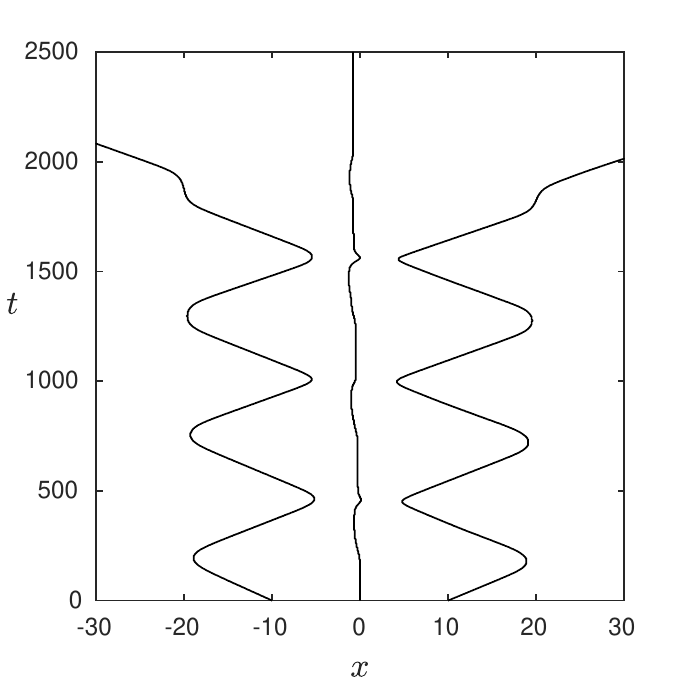}
	\includegraphics[scale =1]{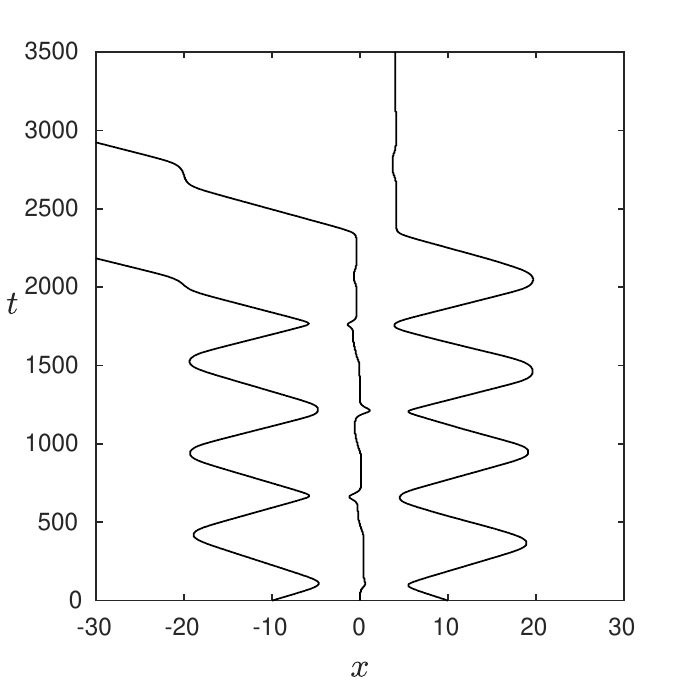}
	\caption{Crest trajectory  for the initial condition  $ \zeta_A(x) + \zeta_B(x) + \zeta_{s}(x)$, where $\zeta_{s}$ is a stationary solution stationed at $x=0$.} 
	\label{colisaoTres}
\end{figure}

Lastly,  we present a remarkable feature when three well separated solitary waves are set as initial condition of equation (\ref{fKdV}).  As can be seen in  Figure  \ref{colisaoTres} (left), one wave remains steady despite the collisions of the other ones. The same goes to simulation in  Figure  \ref{colisaoTres} (right), however a phase shift is noticed.

\section{Conclusion}
In this paper we have studied  solitary trapped waves and collisions for the fKdV equation. We show numerically that for certain regimes those waves  move back and forth oscillating between two bumps until reach a threshold of momentum to escape out. Besides, our experiments indicate that trapped waves have a tendency to escape out upstream.  Regarding collisions of trapped waves, we have shown that  although one wave affects the other,  when we look at the problem statistically the waves behave as they were almost independent. In addition,  when a third steady wave is inserted in the middle of the other ones, we notice that  it remains steady after a short series of collisions  which shows a certain type of stability. 



\section{Acknowledgements}

The authors are grateful to IMPA-National Institute of Pure and Applied Mathematics for the research support provided during the Summer Program of 2020.  M.F. is grateful to Federal University of Paran{\' a} for the visit to the Department of Mathematics. R.R.-Jr is grateful to University of Bath for the extended visit to the Department of Mathematical Sciences.

\bibliographystyle{abbrv}

\end{document}